On the stability of unstable states, bifurcations, chaos of nonlinear dynamical systems. (Shironosov V. G. DAN USSR, 1990, v. 314, No. 2, pp.316-320)

## **ACADEMY OF SCIENCES OF THE USSR**

# REPORT OF THE ACADEMY OF SCIENCES OF THE USSR

# 1990

## VOLUME 314 No.2

### SEPARATE PRINT

# © V.G. SHIRONOSOV
## On the stability of unstable states, bifurcation, chaos of nonlinear dynamic systems

**(Presented by Academician A.N. Tikhonov 24 X 1989)**

The question of the stability of unstable states of dynamical systems that do not explicitly contain a small parameter, chaos and bifurcations in them has attracted attention ever since [1-14]. This is due to the fact that this problem often arises not only in mathematics, but also in various fields of mechanics and physics. In particular, the task of retaining atomic particles in electrodynamic traps has recently become of special interest [14].

As a rule, the solution of such problems is reduced to the study of the model equation pendulum equations with a vibrating suspension point

(1)   $x'' + \varepsilon_r x' + (\varepsilon_o + \varepsilon_1 \cos\tau)\sin x - \varepsilon_{-1}\cos(\tau + \varphi)\cos x = 0,$

At small deflection angles $x$ and $\varepsilon_{-1} = 0$, the equation (1) is reduced to the well-known Mathieu equation, which admits stability of the unstable state of the inverted pendulum ($\varepsilon_o < 0$, $\varepsilon_{-1} \neq 0$) outside the parametric resonance zone [2]. In 1982, the authors [6] found, on the basis of numerical simulation, stable parametrically excited oscillations of the inverted pendulum in the resonance zone. Later [1, 4], the corresponding dependences of the oscillation amplitudes of $\varepsilon_o$, $\varepsilon_1$ were obtained.

In addition to these, many other non-trivial solutions, such as vibrational, vibrational-rotational, the emergence of chaos, etc. [13] were considered. However, the large variety of methods [1-6] and the study (1) with the expansion of sin (x), cos (x) in a series according to a degree of smallness made it difficult to cross-link particular solutions, interpret the results and understand the causes of chaos, bifurcations in systems described by equations of the type (1).

Therefore, taking into account Poincare's two propositions [7, p. 75] that "... the periodic decisions are the only breach through which we could try to penetrate into the region considered to be inaccessible "(1) and that " ... the periodic solution can disappear, only merging with another periodic solution ", i.e. "... the periodic solutions disappear in pairs like the real roots of algebraic equations" (11), we use a generalization of the corresponding methods to find and examine for the stability of periodic solutions (1) with respect to the critical points of the action function [7-12].

To do this, we rewrite equation (1) in the Lagrangian form

(2)   $d(\partial L/\partial x)/d\tau - \partial L/\partial x = -\partial F/\partial x,$

wherein

(3)   $L = T - U, \quad T = x'^2/2, \quad F = \varepsilon_r x'^2/2,$

(4)   $U = -(\varepsilon_0 + \varepsilon_1 \cos\tau)\cos x - \varepsilon_{-1}\cos(\tau + \varphi)\sin x.$

In the general case, $x$ can be a vector. We seek a solution of (2) close to a periodic solution at $\alpha$ frequency in the form of a series

(5)   $x = x_0 + \sum_{n=1}^{\infty} [x_n \cos(n\alpha\tau) + (y_n/n\alpha)\sin(n\alpha\tau)],$

wherein $x_0$, $x_n$, $y_n$ in the general case $f(\tau)$.

Taking into account the dependence $x = f(x_k, y_k, x_k', y_k')$, one can obtain in the approximation of slowly varying amplitudes $x_k$, $y_k$ for the period $2\pi/\alpha$ the following abridged equations:

(6) $x_k' \cong -\partial S/\partial y_k - \partial R/\partial x_k$,   $y_k' \cong \partial S/\partial x_k - \partial R/\partial y_k$,

wherein

$y_k = x_0'$, $k=1, 2, \ldots, \infty$, and

(7) $S = s - y_0^2$,   $s = <L> = (\alpha/2\pi) \int_0^{2\pi/\alpha} L d\tau$,

$$R = (\varepsilon_0/2) \left[ y_0^2 + (1/2) \sum_{n=0}^{\infty} [x_n^2 + y_n^2] \right].$$

In the conclusion (6), the condition of the extremeness of the action function (2) is taken into account. In the variables amplitude–phase, the equations (6) take the following form:

(8) $\psi_n' \cong (1/nr_n) \partial S/\partial r_n$,   $r_n' \cong -(1/nr_n) \partial S/\partial \psi_n - \varepsilon_r r_n$,

$$x = x_0 + \sum_{n=1}^{\infty} [r_n \cos(n\alpha\tau - \psi_n)].$$

In the variables action–angle

(9) $\psi_n' \cong \partial S/\partial \chi_n$,   $\chi_n' \cong -\partial S/\partial \psi_n - 2\varepsilon_r \chi_n$,

$$x = x_0 + \sum_{n=1}^{\infty} [(2\chi_n/n)^{1/2} \cos(n\alpha\tau - \psi_n)].$$

Returning to equation (1), we will seek a solution in the form (8). Using the representation $\cos x = \text{Re}[\exp(ix)]$, formulas (8) and [15]

(10) $\exp[ir_n \cos(n\alpha\tau - \psi_n)] = \sum_{K=-\infty}^{+\infty} Jk_n(r_n) \exp[ik_n(n\alpha\tau + \pi/2 - \psi_n)]$,

we get

(11) $S = \sum_{n=1}^{\infty} n^2 \alpha^2 r_n^2/4 - y_0^2/2 + (1/2) \cdot \sum_{k_1,k_2,\ldots=-\infty}^{+\infty} \prod_{n=1}^{+\infty} Jk_n(r_n) \cdot \sum_{\beta=-1}^{+1} \varepsilon_\beta \delta^{\pm\beta}_{\sum_{N=1}^{\infty} k_n n\alpha}$ (1+

$+ \delta^0_\beta) \cdot \cos\left[ x_0 + \sum_{n=1}^{\infty} k_n(\pi/2 - \delta^{\pm 1}_\beta \psi_n) - \delta^{-1}_\beta (\pi/2 \pm \varphi) \right]$,

wherein $Jk_n(r_n)$ are Bessel functions, and $\delta^n_\beta$ is the Kronecker symbol.

The search for periodic solutions of equations of the type (1), as can be seen from (6), (8), (9) with $\varepsilon_r \cong 0$, is reduced to finding and examining the stability of the critical points (11) with respect to $r_n$, $\psi_n$, or $\chi_n$, $\psi_n$ ($x_n$, $y_n$) and $x_0$, $y_0$.

In the simplest case of a mathematical pendulum, without taking friction and vibrations into account, the results of calculations (8) with respect to $S$ (11) with $n = 1$

(12) $\quad S \cong [\alpha^2 r_1^2/4 + y_0^2/2 + \varepsilon_0 J_0(r_1) \cos x_0]$,

indicate a completely satisfactory accuracy of $\alpha(r_1)$, since the series (11) decreases rapidly with increasing index $n$ for a fixed value of the argument $r_n$.
A relative error of the approximation $\alpha(r_1)$ even for angles of deviation of the pendulum $x \sim 160°$ does not exceed 5.5% (see, for example, [5, p. 55]).

The introduction of longitudinal vibration, as follows from expression

(13) $\quad S \cong [\alpha^2 r_1^2/4 - y_0^2/2 + \varepsilon_0 J_0(r_1) \cos x_0 + \varepsilon_1 J_{1/\alpha}(r_1) \cos(x_0 + \pi/2\alpha) \cos(\psi_1/\alpha)]$,

and (8), results in two types of critical points. The first ones are the equilibrium positions $x_0 = \pm n\pi$, $\psi_1 = 0, \pm \pi/2$, ($1/\alpha$ are even); the second ones are $x_0 \neq \pm n\pi$, $\psi_1 = 0, \pm \pi/$, ($1/\alpha$ are uneven), $n = 0, 1, 2, \ldots$ (in particular, $x_0 = \pm (2n+1)$ for $\varepsilon_0 = 0$). Therefore, taking into account the scenario of "merging" of two periodic solutions according to Poincare (11), due to the presence of the second type of critical points $x_0 \neq \pm n\pi$ (bifurcation of the period $1/\alpha = 2 \leftrightarrow 1/\alpha = 1$), we seek the solution for the problem of an inverted pendulum ($\varepsilon_0 \cos x_0 < 0$) outside and in the zone of parametric resonance in the following form:

(14) $\quad x = x_0 + r_1 \cos(\tau/2 - \psi_1) + r_2 \cos(\tau/2 - \psi_2)$.

This representation of (14) leads to the expression $S$ (11) with an accuracy of $n = 2$. When limiting to the terms of the order $r_k^4$ with the expansion of $J_n(r_k)$ in $S$ (11) and using the variables $x_k$, $y_k$ (6), we can obtain the corresponding equations to find equilibrium points and the characteristic roots $\lambda_0$ for small $\varepsilon_r$, in the analytic form.
In the case $x_1 = x_2 = y_1 = y_2 = \sin x_0 = y_0 = 0$

(15) $\quad \{\lambda^2 + 1/16[(1 - 4\varepsilon_0^\pm)^2 - 4(\varepsilon_1^\pm)^2]\}\{\lambda^4 + \lambda^2(1 + \varepsilon_0^\pm)^2/4 + 1/8(1 - \varepsilon_0^\pm)[(\varepsilon_1^\pm)^2 + 2\varepsilon_0^\pm(1 - \varepsilon_0^\pm)]\}$.

wherein $\lambda = \lambda_0 + \varepsilon_r$, $\varepsilon_{0,1}^\pm = \varepsilon_{0,1} \cos x_0$. From the first bracket (15) we obtain the upper limit evaluation of the stable solution $4(\varepsilon_1^\pm)^2 < (1 - 4\varepsilon_0^\pm)^2$, from the second we obtain the lower one $(\varepsilon_1^\pm)^2 > 2|\varepsilon_0^\pm(1 - \varepsilon_0^\pm)|$ for an inverted pendulum ($\varepsilon_0^\pm < 0$) outside the parametric resonance.

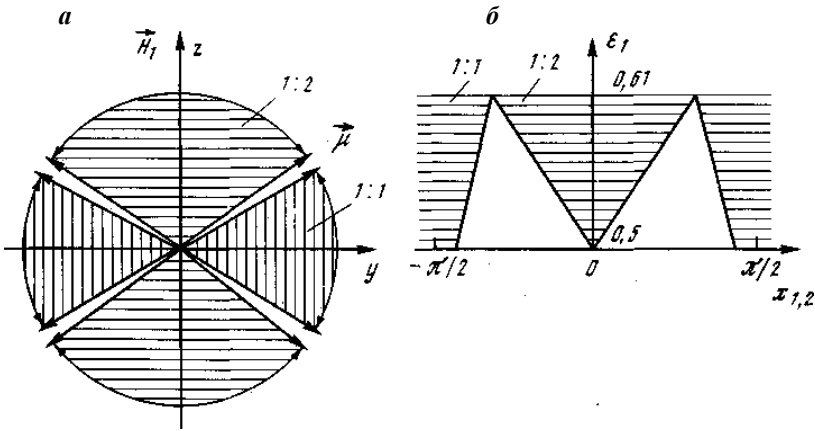

а                          б

Fig. 1. Scenario for the appearance of a bifurcation for an inverted pendulum according to Poincare with $\varepsilon_0=0$. a) $0,5<\varepsilon_1<0,61$; dependency graphs of $x_{1,2}(\varepsilon_1)$.

In the case $x_1 \neq 0$, $y_1 = 0$ ($x_1 = 0$, $y_1 \neq 0$), from the conditions $\partial S/\partial x_1 = 0$ ($\partial S/\partial y_1 = 0$), we can get

(16) $x_1^2 = 6[(4\varepsilon_0^\pm + 2\varepsilon_1^\pm - 1)/(2\varepsilon_1^\pm + 3\varepsilon_0^\pm)]$, ( $y_1^2 = (3/2)(4\varepsilon_0^\pm - 2\varepsilon_1^\pm - 1)/(3\varepsilon_0^\pm - 2\varepsilon_1^\pm)]$ ),

(17) $[\lambda^2 + (x_1^2/24)[2\varepsilon_0^\pm + \varepsilon_1^\pm + 1]]f_x(\lambda) = 0$, ($[\lambda^2 - \varepsilon_1^\pm y_1^2(2\varepsilon_0^\pm - 2\varepsilon_1^\pm - 1)/6]f_y(\lambda) = 0$ ),

wherein $f_y(\lambda) = f(\lambda, \varepsilon_{0,1}^\pm, x_1, y_1)$. It results from (17) that there are two stable states of the inverted pendulum ($\varepsilon_0^\pm < 0$) in the zone of parametric resonance $2\varepsilon_1^\pm > 4|\varepsilon_0^\pm| + 1$, ($2|\varepsilon_1^\pm| > 4|\varepsilon_0^\pm| + 1$), differing from each other only in the change of sign in $\varepsilon_1^\pm$.

In the simplest case with $\varepsilon_0 = 0$, the bifurcation point is $1/\alpha = 2 \leftrightarrow 1/\alpha = 1$ is found from the joint consideration of two periodic solutions under the scenario (11). Carrying out similar calculations near the equilibrium point $x_0 = \pm (2n+1)\pi/2$, $x_1 = y_1 = y_0 = 0$, one can obtain solutions with $\alpha^1 = 1$

(18)   $x_2 \cdot 4(1-(1+3\varepsilon_1^{*2}/2)^{1/2})/3\varepsilon_1^*$,   $\varepsilon_1^* = \varepsilon_1 \sin x_0$, $y_2 = 0$,

are unstable with respect to $x_0$, $y_0$ for $|x_2| = \pi/2$. Solving jointly (16), (18), one can define the corresponding bifurcation point from the condition (see Fig. 1):

(19)   $|x_1^*(\varepsilon_1 N)| + |x_2^*(\varepsilon_1 N)| = \pi/2$,

$x_1^* . 59°$,  $x_2^* . 31°$,  $\varepsilon_1 N . 0.61$.

In this case (with $\varepsilon_0 = 0$), the appearance of a bifurcation can simultaneously lead to the emergence of chaos in the system (1) (see Fig. 1). The reason for this may be fluctuations, errors from the macrosystem used in the physical, analog or numerical simulation of the deterministic system described by the equation (1). As a result, cascades of transitions between different types of periodic motions $\varepsilon_1 = \varepsilon_1 N$ (vibrational 1:2, 1:1, rotational 1:1 and others), perceived as chaos, will be observed.

Computer simulation of the equation (1) using the analog computing machinery "Rusalka" and full-scale modeling with a magnetic needle of a compass placed in a magnetic field confirmed the correctness of the results obtained within the limits of simulation errors.

The author is grateful to S.P. Kurdyumov, Yu.P. Popov, Sarychev V.A. and other participants of the seminar for discussion of the work and useful comments.

Physics and Technology Institute
the Ural Branch of the Academy
of Sciences of the USSR
Izhevsk

Translated by Shironosova O. E.
shironosova.pr@gmail.com